\documentclass[fleqn,10pt]{wlscirep}
\usepackage[utf8]{inputenc}
\usepackage[T1]{fontenc}
\usepackage{lineno}
\usepackage{physics}
\usepackage{comment}
\usepackage{array}
\usepackage{booktabs}
\usepackage{caption}
\usepackage{subcaption}
\usepackage{graphicx}

\title{Improving Free-Space Continuous Variable Quantum Key Distribution with Adaptive Optics}

\author[1,2,3,4,5*]{Mikhael T. Sayat}
\author[6,7]{Marcus Birch}
\author[7]{Michael Copeland}
\author[7]{Elisa Jager}
\author[8,5]{Oliver Thearle}
\author[7]{Francis Bennet}
\author[1,2,3,5]{Ping Koy Lam}
\author[4]{Nicholas J. Rattenbury}
\author[9]{John E. Cater}

\affil[1]{Quantum Innovation Centre (Q.InC), Agency for Science Technology and Research (A*STAR), 2 Fusionopolis Way, Innovis \#08-03, Singapore 138634, Republic of Singapore}
\affil[2] {Institute of Materials Research and Engineering (IMRE), Agency for Science Technology and Research (A*STAR), 2 Fusionopolis Way, Innovis \#08-03, Singapore 138634, Republic of Singapore}
\affil[3] {Centre for Quantum Technologies, National University of Singapore, 3 Science Drive 2, Singapore 117543, Singapore}
\affil[4]{Department of Physics, Faculty of Science, University of Auckland, Auckland 1010, New Zealand}
\affil[5]{Centre of Excellence for Quantum Computation and Communication Technology,
The Department of Quantum Science and Technology, Research School of Physics and Engineering,
The Australian National University, Canberra, Australian Capital Territory 2601, Australia
}
\affil[6]{Swedish Space Corporation, Solna 17154, Sweden}
\affil[7]{Advanced Instrumentation Technology Centre, Research School of Astronomy and Astrophysics, Australian National University, Canberra, 2601, Australia}
\affil[8]{Quantum Technology Group, Defence Science \& Technology Group, Adelaide 5111, Australia}
\affil[9]{Department of Mechanical Engineering, University of Canterbury, Christchurch 8041, New Zealand}
\affil[*]{Corresponding author: mikhael\_sayat@imre.a-star.edu.sg}


\begin{abstract}
A significant performance inhibitor of free-space continuous variable quantum key distribution (CVQKD) is turbulence, which gives rise to wavefront phase and amplitude aberrations. We demonstrate that in a turbulent channel, during coherent state transmissions from a continuous-wave laser, that the interferometric visibility between the local oscillator (LO) and quantum signal decreases. A solution to this is incorporating adaptive optics at the receiver to correct phase and amplitude aberrations in the wavefronts of the received quantum signal. We demonstrate the increased interferometric visibility and decrease in its fluctuations in a 60 cm and 30 m turbulent channel when using adaptive optics through channel characterisation. In an ideal CVQKD system, we show that this leads to more precise and larger positive secret key rates, improving the performance of free-space CVQKD in turbulent channels.  
\end{abstract}
\begin{document}

\flushbottom
\maketitle
%
%
\thispagestyle{empty}

\section{Introduction}

Continuous variable quantum key distribution (CVQKD) experiments have predominantly been restricted to fibre-based implementations with the first 100 km fibre transmission reported in 2016 \cite{Huang2016Long}. Multiplexing has since been employed to transmit more information through fibre, achieving larger data rates \cite{Qu2017High, Qu2018High}. Unfortunately, transmission through a turbulent free-space channel has been limited to feasibility studies and simulations \cite{kish2020feasibility, sayat2024satellite, sayat2025dynamic}, experimental demonstrations with simulated light modulators \cite{Qu2017Approaching}, and proposed complete systems \cite{Wei2023High, Zhang2023Experimental}. In fact, CVQKD has only been demonstrated over a 460~m channel, which used a Gaussian distribution to modulate the polarising quantum Stokes parameter unidimensionally \cite{Shen2019Free}, as well as over 860~m using Gaussian modulated coherent state CVQKD \cite{zheng2025free}.


Turbulence is known to have detrimental effects on a free-space optical communications channel. Variations in the refractive index of the medium distort optical wavefronts giving rise to phase and amplitude aberrations (scintillation), \cite{Bennet2020Australia} and beam wandering \cite{Berman2007Beam, Kaushal2011Experimental}. Designs of CVQKD systems that have incorporated adaptive optics (AO) have been proposed to improve the performance of CVQKD in turbulent free-space channels by suppressing noise, from the channel and the use of AO, and therefore increasing the secret key rate (SKR) \cite{Bennet2018Free, Wang2019Performance, Chai2020Suppressing}. Feasibility studies, simulations, and analyses of CVQKD with AO in a satellite-to-ground link have also been performed \cite{Villasenor2020Atmospheric, Acosta2021Analysis, Acosta2023Improvement}. Despite the improvements from AO that lead to increased SKRs in turbulent channels through simulations, there has yet to be an implementation of CVQKD with AO to experimentally validate this claim.


In this work, we experimentally study the effects of turbulence on coherent state transmission in free-space for CVQKD using a continuous-wave laser through channel characterisation. We quantify the effects of turbulence on the interferometric visibility between a quantum signal beam and local oscillator (LO) during detection \cite{Lam1998Applications}, the consequential impact on the measured variance of the coherent state \cite{Roumestan2022Experimental}, and its impact on the secret key rate (SKR). An AO system has been implemented at the receive terminal as a method to correct the wavefront aberrations. The corrected wavefronts resulted in improved mode-matching between the quantum signal and LO, leading to improved interferometric visibility. Experiments in a 60 cm and 30 m channel have been performed in free-space where turbulence was introduced using a heat gun. The turbulence is characterised by the variance of the movement of the deformable mirrors in the AO system. The measured interferometric visibility was then used to calculate the SKR for CVQKD.

\section{Background}

The baseline experimental setup and overall methodology in this section is based on previous work (Ref. \citenum{sayat2024experimental}) which has been extended to the 30~m link.

\subsection{Interferometric Visibility Measurement}
\label{VM}
In fibre implementations, the quantum signal and local oscillator (LO) are both contained in fibre, and are therefore mode-matched with an interferometric visibility of $\sqrt{\eta_{\mathrm{vis}}} = 1$. However, in free-space CVQKD, this is not the case. The quantum signal is degraded in free-space and is not guaranteed to be mode-matched with the LO. The interferometric visibility, therefore, plays a significant role in free-space CVQKD as it can be used to infer loss that can be used to calculate the SKR.

The interferometric visibility, henceforth referred to as the "visibility", between the quantum signal and LO used manifests as interference fringes on the oscilloscope. This comes from the direct current (DC) signal of the homodyne detector as the phase between the quantum signal and LO is scanned \cite{Lam1998Applications}. The visibility is calculated as
\begin{equation}
    \sqrt{\eta_{\mathrm{vis}}} = \frac{I_{\mathrm{max}} - I_{\mathrm{min}}} {I_{\mathrm{max}} + I_{\mathrm{min}}
    \label{Vis}},
\end{equation}
where $I_{\mathrm{max}}$ and $I_{\mathrm{min}}$ are the maximum and minimum value of the fringes on the oscilloscope.

Before coherent state measurements, when there is only ambient turbulence (no turbulence introduced by a heat gun), baseline measurements of the visibility were performed and calculated as in Equation \ref{Vis} using the DC signal from a homodyne detector. However, during coherent state measurements, the phase between the signal and LO is locked such that the DC signal is zero, making the visibility impossible to measure using Equation \ref{Vis}. Therefore, the inferred visibility is taken as the reduced power of a phase modulated signal relative to a baseline measurement. Since the DC signal is used for phase-locking, the inferred visibility is taken as the mean of the measured coherent state trace normalised to the visibility during baseline measurements.

\subsection{SKR Calculation}

\label{sec:SKRC}
The calculation of the SKR assumes Gaussian modulation with security under collective attacks in the asymptotic limit \cite{Denys2021Explicit}. Further details can be found in Appendix \ref{AppendixA}. The SKR is defined as

\begin{equation}
    \mathrm{SKR} = \beta I_{AB} - S_{BE},
    \label{eqn:SKR_asy}
\end{equation}
where $\beta$ is the reconciliation efficiency, and $I_{AB}$ is the mutual information between the transmitter, Alice, and receiver, Bob. $S_{BE}$ is the estimated amount of information an eavesdropper, Eve, knows (Holevo bound). 

The covariance matrix is defined as 

\begin{equation}
    \Gamma = 
    \begin{bmatrix}
    (V_A + 1) \mathbf{I}  &  Z\mathbf{\sigma_z} \\
    Z\mathbf{\sigma_z}  &  V_B\mathbf{I}
    \end{bmatrix} = 
    \begin{bmatrix}
    V\mathbf{I}  &  Z\mathbf{\sigma_z} \\
    Z\mathbf{\sigma_z}  &  V_B\mathbf{I}
    \end{bmatrix},
    \label{CM}
\end{equation}
where $V_A$ is the modulation variance which is related to the modulation amplitude, $V_A = 2\alpha^2$, $Z$ is the correlation coefficient, and $V_B$ is Bob's variance. A significant factor in defining the covariance matrix is the inclusion of the visibility, $\sqrt{\eta_{\mathrm{vis}}}$, in $Z$ and $V_B$. All instances of the transmittance, $T$, is then multiplied by $\eta_{\mathrm{vis}}$ in the calculation of the SKR.

The correlation coefficient for Gaussian modulation is therefore calculated as

\begin{equation}
    Z = 2\sqrt{\eta_{\mathrm{vis}}\eta_{\mathrm{det}}T}\sqrt{\alpha^4 + \alpha^2},
\end{equation}
where $T$ is the transmittance, $\eta_{\mathrm{det}}$ is the detector efficiency, and $\alpha$ is the modulation amplitude. $T$ is the ratio between the received power, $P_R$, and the transmitted power, $P_T$, with ambient turbulence (no heat gun) i.e. $T = \frac{P_R}{P_T}$. 

The measured variance at Bob can be calculated depending on the CVQKD framework \cite{Pirandola2021Composable}. Here, Bob's modelled variance is calculated as 

\begin{equation}
    \label{VB}
    V_{B} = 1 + \eta_{\mathrm{vis}}\eta_{\mathrm{det}}T(2\alpha^2 + \xi) + v_{\mathrm{el}}.
\end{equation}





\subsection{Implementation}
\label{ExperimentSetup}

\begin{figure}[htp!]
    \centering    \includegraphics[width=0.55\textwidth]{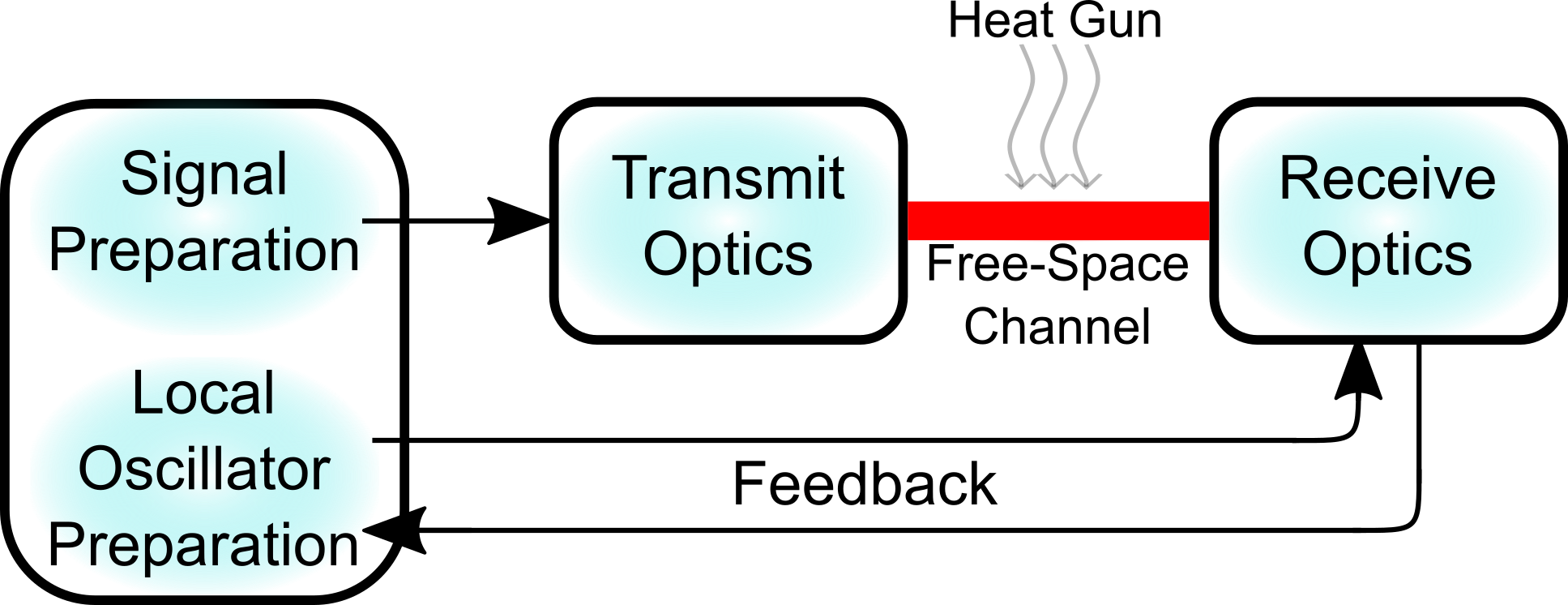}
    \caption{Block diagram of the experimental setup between transmit (Alice) and receive (Bob) optics through a free-space channel with introduced turbulence from a heat gun. A closed-loop feedback is used between Receive Optics and Local Oscillator Preparation to control the phase relation between the quantum signal and local oscillator (LO).}
    \label{fig:ExperimentalSetup}
\end{figure}

A schematic of the general experimental setup is shown in Figure \ref{fig:ExperimentalSetup}, where coherent states are prepared in fibre and transmitted optically from Alice to Bob through a 60 cm and 30 m free-space channel, separately. Turbulence is introduced in the free-space channel using a heat gun with varying power settings. A closed-loop feedback signal is used to control the phase relation between the quantum signal and the LO at the receiver, ensuring a phase-lock.

\begin{figure}[htp!]
    \centering
    \includegraphics[width=0.75\textwidth]{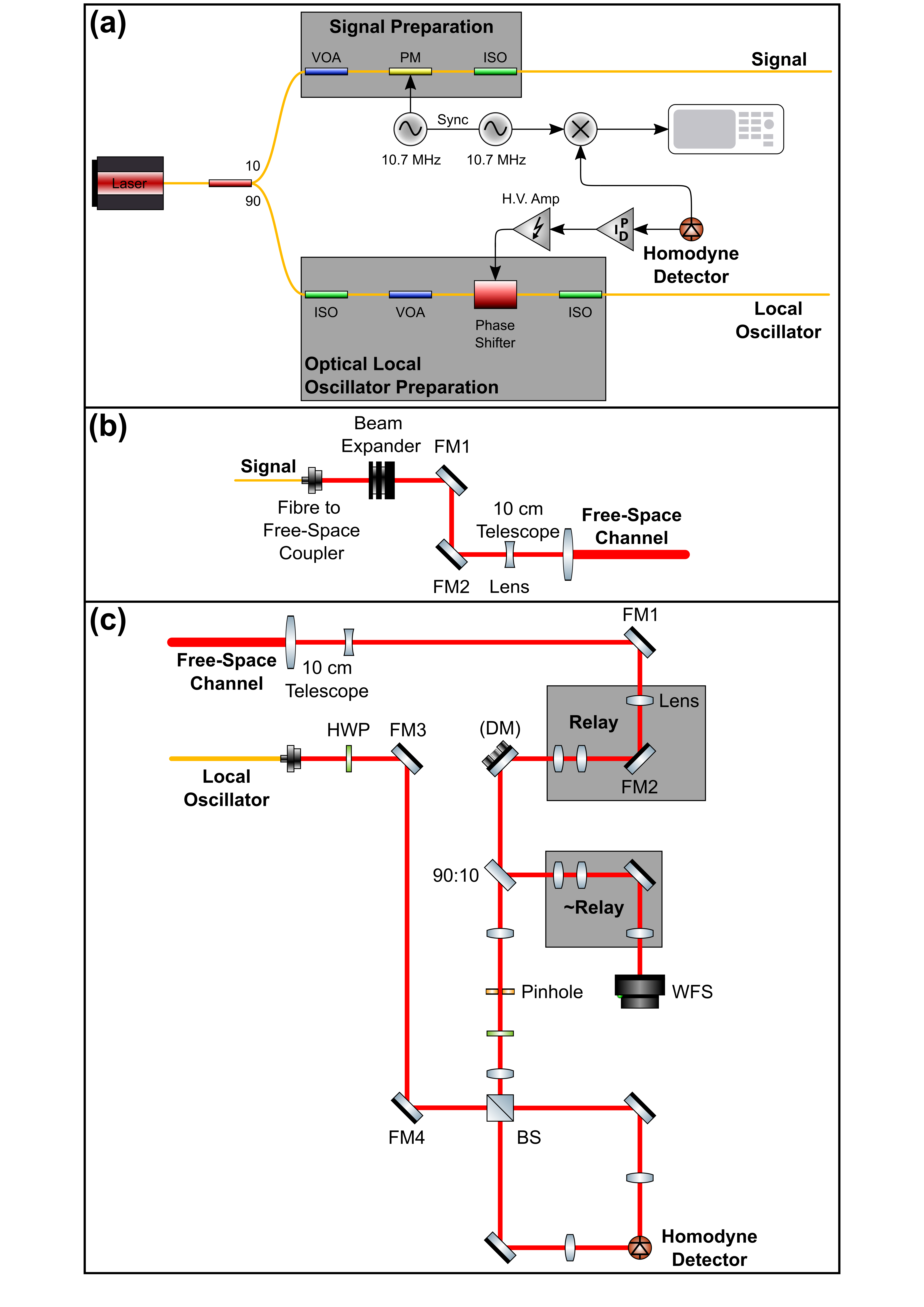}
    \caption{(a) State Preparation and Local Oscillator Preparation in fibre. The laser is split into the quantum signal beam and the local oscillator (LO) beam. Feedback is based on the received homodyne detection to ensure a phase-lock between the LO and the quantum signal. (b) Transmit Optics. The quantum signal beam from State Preparation is coupled to free-space and transmitted through a free-space channel where turbulence is introduced through varying power settings of a heat gun (not shown). Mirrors, FM1 and FM2, are used for alignment. Beam expander used to mitigate beam divergence in the free-space channel. (c) Receive Optics. The quantum signal beam is received and split for turbulence characterisation and interference with the LO for homodyne detection. The LO is coupled to free-space from State Preparation. An adaptive optics (AO) system is integrated through a closed-loop feedback between the deformable mirror (DM) and wavefront sensor (WFS). Note: The homodyne detector in (a) and (c) is the same. Yellow lines indicate fibre connection. Red lines indicate free-space connections.}
    \label{fig:Together}
\end{figure}

Figure \ref{fig:Together} shows a schematic for coherent state generation and detection. A 10 mW 1550 nm continuous-wave fibre laser is split 90:10 for the LO and quantum signal, respectively (Figure \ref{fig:Together}a). The quantum signal power is controlled by a variable optical attenuator (VOA) and modulated using a phase modulator (PM). The modulation used a 10.7 MHz sine wave from a function generator. An isolator (ISO) is used as a polarisation filter to prevent back propagation of light in the fibre. The quantum signal is then coupled to the transmit optics in free-space (Figure \ref{fig:Together}b). Here, it is expanded, to reduce beam divergence, and sent through a 10 cm diameter telescope to the free-space channel towards the receive optics. The mirrors (FM1 and FM2) are used for spatial alignment. The quantum signal is received by a 10 cm diameter telescope and is split 90:10 towards the homodyne detector and a Shack-Hartmann wavefront sensor (WFS), respectively (Figure \ref{fig:Together}c). 

The light on the WFS path is used to characterise turbulence in the free-space channel. The WFS calculates the ``slopes" in its sub-apertures (movement of spots from the received light) which are then used to calculate the slope variance over a measurement period for turbulence characterisation. The light on the homodyne detector path passes through a lens pair to resize the beam with a pinhole at the focal point to clean the spatial mode of the received beam, and a half-wave plate (HWP) for polarisation control. On the WFS beam path, the signal passes through a series of lenses (Relay) to achieve the correct magnification and illumination on the deformable mirror (DM) as well as placing the DM in a conjugate plane to the telescope primary.  The $\sim$Relay has been implemented to remove the effects of the Relay on the WFS. In particular, resizing the beam so that the DM and WFS are conjugated. This is to increase the performance of the AO system which takes images of the beam at the WFS and adjusts the DM actuators accordingly, forming a closed-feedback loop for corrections.



The LO is first passed through an ISO to reduce polarisation drift and a VOA for power control (Figure \ref{fig:Together}a), and is then coupled to free-space (Figure \ref{fig:Together}c). It passes through a HWP for polarisation control and two mirrors (FM3 and FM4) for alignment. The quantum signal and LO interfere at the beam splitter (BS). It is important that the two beams are mode-matched to maximise the visibility between the signal and LO and subsequently the detection efficiency. To ensure that the two beams are mode-matched, they both separately pass through HWPs for polarisation control. In addition, the LO is spatially aligned to the quantum signal using mirrors FM3 and FM4. A lens was used before the BS to resize the quantum signal beam to match the LO. The two arms after the BS are spatially aligned and focused on the photodiodes in the homodyne detector using mirrors and lenses. A second 10.7 MHz sine wave, synced with the 10.7 MHz optical modulation, is used to demodulate the AC signal from the homodyne detector and sent to the oscilloscope for coherent state measurements. 


A phase shifter is used to control the phase relation between the quantum signal and LO during homodyne detection (Figure \ref{fig:Together}a), creating the feedback as shown in Figure \ref{fig:ExperimentalSetup}. The received DC interference signal from the homodyne detector is passed through an optimised PID controller and high voltage amplifer (H.V. Amp) to lock the received LO and quantum signal out of phase, maximising the 10.7 MHz signal at the receiver. This allows the phase quadrature to be measured. Note that this DC signal is used to calculate and optimise the visibility. An ISO is once again used to prevent back propagation of light in the fibre.

Received power, $P_R$, was measured as the power of the quantum signal just before the BS (Figure \ref{fig:Together}c) and the transmitted power, $P_T$, was measured as the quantum signal power just after the fibre-to-free-space coupler (Figure \ref{fig:Together}b). The transmittance, $T = P_R/P_T$, is therefore an overall measurement of transmittance based on the free-space channel and the optical efficiency from transmit optics to receive optics before the BS.




The AO system consists of the DM and WFS (Figure \ref{fig:Together}c) and has a closed-loop bandwidth of 135 Hz. The signal beam received on the WFS path is split into 36 sub-apertures via a lenslet array and forms a feedback loop with the DM. The DM actuators, which have an operating frequency of 2 kHz, are displaced from their initial position to correct the distorted quantum signal beam as captured and measured on the WFS when turbulence is present. Registration (mapping of DM actuator movements to WFS sub-apertures) is ensured by calculating an interaction matrix where the movement of each DM actuator has an acceptable response as captured by the WFS \cite{bennet2024NGTF}. The interaction matrix is then used to build a reconstructor, which relates the slopes measured by the WFS to the appropriate command for the DM actuators.

\subsection{Coherent State Measurement}
\label{CSM}

The coherent states measured by the homodyne detector are recorded on the oscilloscope and normalised to the shot noise. This requires three different traces on the oscilloscope to be recorded: the coherent state trace, the shot noise trace, and the dark noise trace. The traces were recorded with 1 million data points over 240 ms. The traces were measured with the following experimental settings:
\begin{itemize}
    \item Coherent state trace: Quantum signal and LO on the homodyne detector.
    \item Shot noise trace: Only the LO on the homodyne detector. No quantum signal (blocked).
    \item Dark noise trace: No quantum signal and LO on the homodyne detector, i.e. laser is off.
\end{itemize}
The coherent state trace was then sectioned into 20 windows corresponding to 50,000 data points over 12 ms, and normalised to the true shot noise (shot noise without dark noise). This gives 20 coherent state measurements per coherent state trace, each acting as an individual normalised coherent state measurement.

Bob's variance, normalised to shot noise (SNU -- shot noise units), is calculated as \cite{Laudenbach2018Continuous}
\begin{equation}
    V_{B} = \frac{CS}{SN - DN} \; [\mathrm{SNU}],
\end{equation}
where $CS$ is the variance of the coherent state trace, $SN$ is the variance of the shot noise trace, and $DN$ is the variance of the dark noise trace. The associated dark noise (or electronic noise) is calculated as
\begin{equation}
    v_{\mathrm{el}} = \frac{DN}{SN - DN} \; [\mathrm{SNU}].
    \label{eqn:electronicNoise}
\end{equation}
Note that $DN$ is subtracted from $SN$ to remove any noise outside the LO, normalising the coherent states to the true shot noise. As the measured electronic noise is constant at $v_{\mathrm{el}}=0.027$ (SNU), Bob's electronic noise was trusted in the experiment and subtracted from measurements. 
The excess noise is calculated as

\begin{equation}
    \xi = \frac{\mu(V_B - V_{\mathrm{el}} - 1)}{T} \; [\mathrm{SNU}].,
    \label{eqn:excessnoise}
\end{equation}
where $\mu = 1$ for homodyne detection and $\mu = 2$ for heterodyne detection.


\newpage
\subsection{Turbulence Characterisation}
\label{ScintillationM}
\begin{figure}[htp!]
    \centering
    \includegraphics[width=0.55\textwidth]{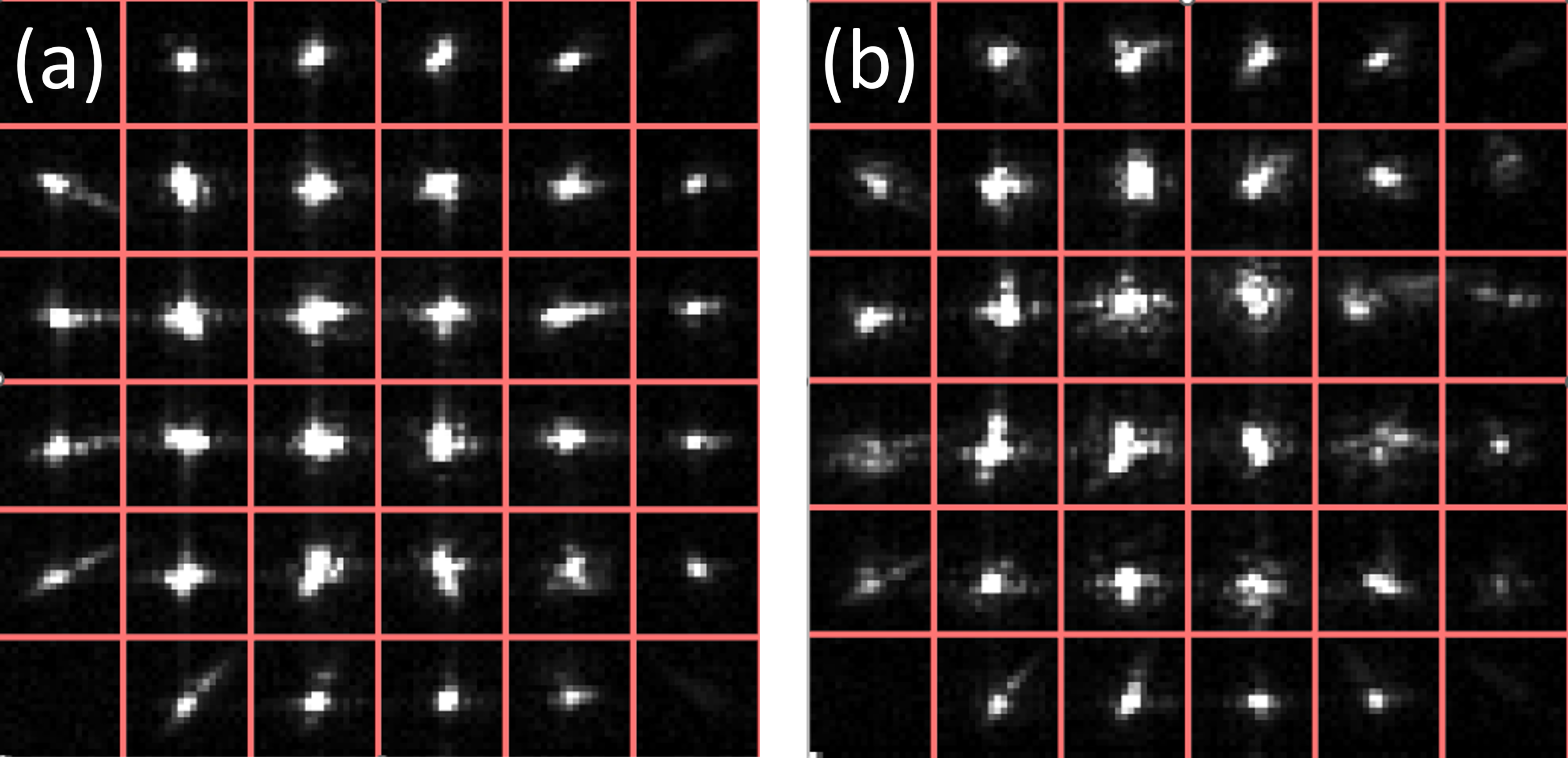}
    \hfill
    \caption{Quantum signal beam on the Shack-Hartmann WFS (a) without and (b) with turbulence introduced by the heat gun. The AO system corrects the wavefront aberrations in (b) to (a) through a closed-loop feedback between the received images in the WFS and the appropriate commands on the DM actuators.}
    \label{fig:WFS}
\end{figure}

Figure \ref{fig:WFS} shows the typical behaviour of the quantum signal beam on the WFS sub-apertures with and without the turbulence introduced by the heat gun. It can be seen that the beam undergoes scintillation causing intensity fluctuations in each sub-aperture. The AO system tries to correct the distorted wavefront from the turbulence introduced by the heat gun back to the original wavefront when there is no turbulence from the heat gun. The appropriate commands are sent to the DM actuators to correct the wavefront based on the images on the WFS.



The inner 32 sub-apertures of the 36 sub-apertures (corner sub-apertures excluded due to insignificant intensity) in the WFS were used to register commands to the DM actuators to correct the scintillated beam back to its original form (when there is no turbulence from the heat gun) through a closed-loop feedback. The centre of each spot in the inner 32-sub-apertures was calculated by the AO software. The displacement or "slope'' of the spot from the centre is then calculated as

\begin{equation}
    \mathrm{slope} = \sqrt{X^2 + Y^2},
\end{equation}
where $X$ and $Y$ are displacements in the $X$ and $Y$ axes. The average slope of a frame (the inner 32 sub-apertures) is then calculated. The variance of the average slope or "slope variance'', calculated using the open-loop feedback of the AO system over a measurement period is then used to characterise the turbulence (12,000 frames with an operating frequency of 2 kHz). Note that the open-loop feedback slope variance was used as this measured the turbulence without the AO system, thus providing a more accurate representation of the turbulence. A large slope variance corresponds to larger disturbances from a high power setting on the heat gun. Conversely, a small slope variance corresponds to smaller disturbances from a low power setting. The slope variance characterises the turbulence relative to each value and it increases with the scintillation index. Unless otherwise stated, measurements of the visibility at a given slope variance have been
made when a phase-lock is maintained.

\section{Results}
The visibility measurements along with the SKR calculations have been analysed for both the 60 cm and 30~m free-space channels. In addition, the noise measured in the coherent states are also analysed and discussed. Further detailed results can be found in Appendix \ref{AppendixB}. 


\subsection{The 60 cm Free-Space Channel}
\begin{figure}[htp!]
    \centering
    \includegraphics[width=0.45\textwidth]{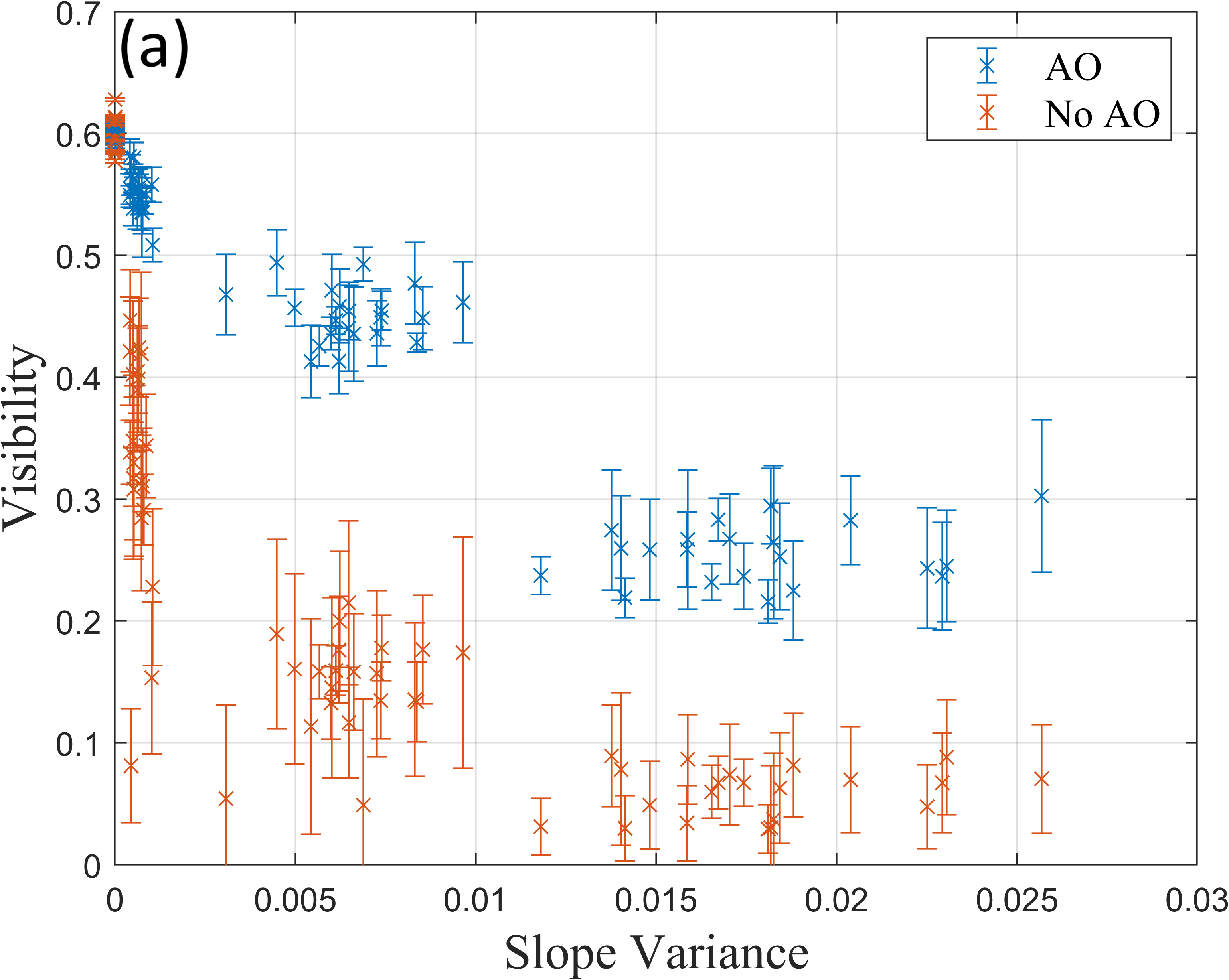}
    \includegraphics[width=0.45\textwidth]{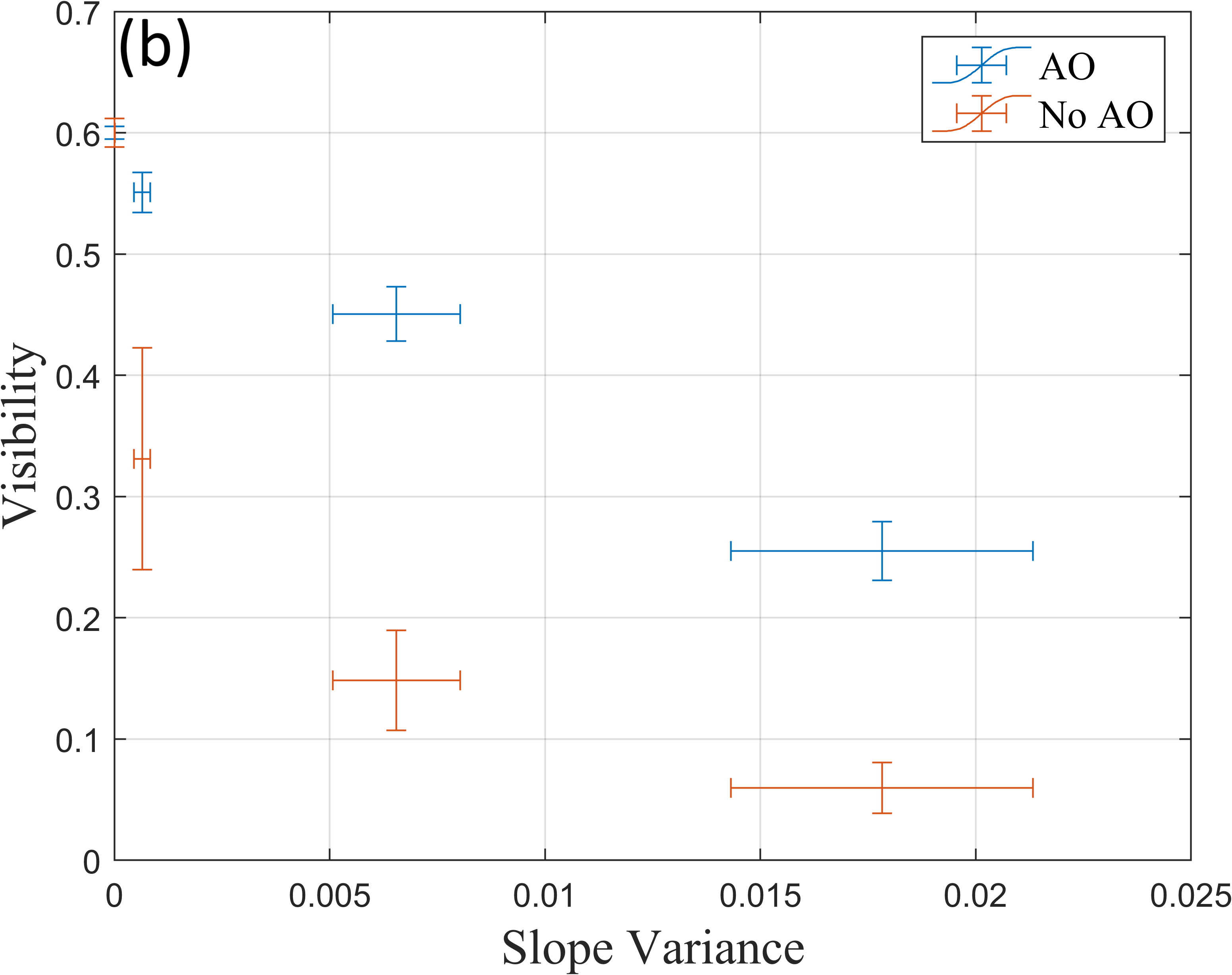}
    \caption{The (a) individual visibility measurements and (b) the mean visibility for each heat gun power setting, as a function of the slope variance both with and without the AO system in the 60 cm free-space channel.}
    \label{fig:Visibility60cm}
\end{figure}

The effects of increasing turbulence in a 60 cm free-space channel are displayed in Figure \ref{fig:Visibility60cm}. It should be noted that only four discrete power settings on the heat gun were used and these are represented as clusters in Figure \ref{fig:Visibility60cm}. Figure \ref{fig:Visibility60cm}b shows the mean visibility and slope variance for each power setting. From left to right: ambient turbulence (no heat gun), low turbulence (low power), medium turbulence (medium power), high turbulence (high power). The system had a visibility of $\sqrt{\eta_{\mathrm{vis}}} = 0.6$ in ambient turbulence.

A significant drop in the visibility was observed with a low heat gun setting (Slope Variance $=6.5 \times 10^{-4} \pm 1.9 \times 10^{-4}$), with smaller decreases as the slope variance increases. When using the AO system, the decrease in the visibility is reduced and there is an overall larger visibility. It can also be seen that the variation in the individual visibility measurements, represented as the standard deviation (error bars in Figure \ref{fig:Visibility60cm}a), decreased by using the AO, except at high heat gun settings where it stays the same (Table \ref{tab:MeanSD}). This could be a sign that the AO system struggles to correct the channel fluctuations at high heat gun settings (slope variance $= 0.018 \pm 0.0035$). Table \ref{tab:MeanSD} also shows that the standard deviation of the visibility increases with more powerful heat gun settings. This may be due to the increase in scintillation, where the beam scatters on the photodiode resulting in larger fluctuations in the visibility. 

\begin{table} [htp!]
\centering
\caption{Mean Standard Deviation in the Visibility of Measured Coherent States}
\scalebox{1}{
\begin{tabular}{|l|c|c|}
    \bottomrule
    Heat Gun Setting & AO & No AO \\ \toprule \bottomrule
    Ambient (no heat gun) & $0.001$ & $0.002$ \\
    \hline
    Low & $0.014$ & $0.049$\\
    \hline
    Medium & $0.024$ & $0.056$ \\
    \hline
    High & $0.037$ & $0.037$ \\
    \toprule
\end{tabular}}
\label{tab:MeanSD}
\end{table}

Figure \ref{fig:Visibility60cm}b shows the mean visibility for each heat gun power setting used with and without the AO system. This allows the calculation of the retrieved visibility at each heat gun power setting and thus a measurement of the performance of the AO system. The difference between the visibility with and without the AO system is shown in Table \ref{tab:MeanInterferometricVisibilities}. The results show that the AO system had the largest difference when the heat gun had a medium power setting.

\begin{table} [htp!]
\centering
\caption{Mean Visibility}
\scalebox{1}{
\begin{tabular}{|l|c|c|c|}
    \bottomrule
    Heat Gun Setting & AO & No AO & Difference \\ \toprule \bottomrule
    Ambient (no heat gun) & $0.60$ & $0.60$ & $-$ \\
    \hline
    Low & $0.55$ & $0.33$ & $0.22$\\
    \hline
    Medium & $0.45$ & $0.15$ & $0.30$ \\
    \hline
    High & $0.26$ & $0.06$ & $0.20$ \\
    \toprule
\end{tabular}}
\label{tab:MeanInterferometricVisibilities}
\end{table}

\begin{figure}[htp!]
    \centering
    \includegraphics[width=0.45\textwidth]{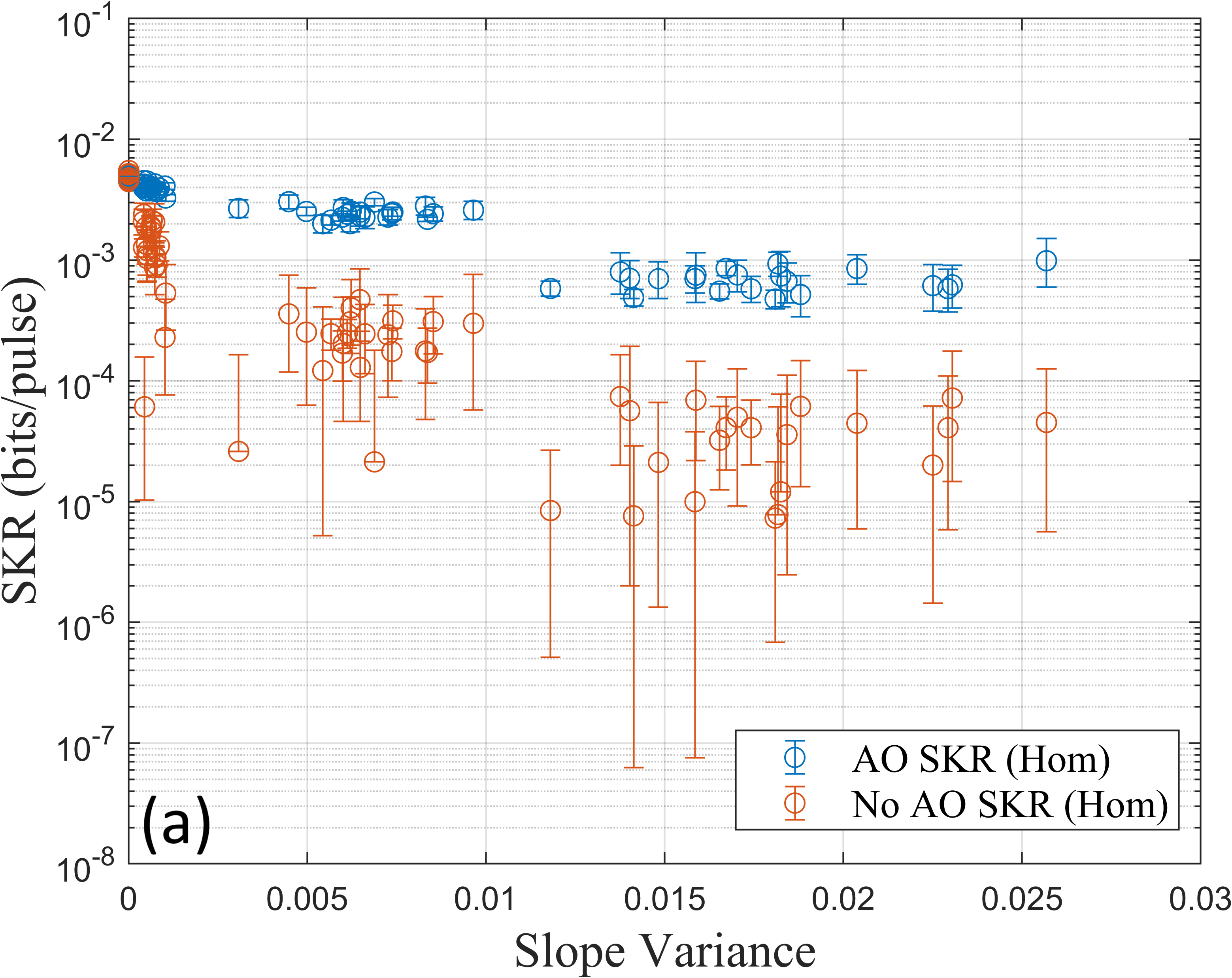}
    \includegraphics[width=0.45\textwidth]{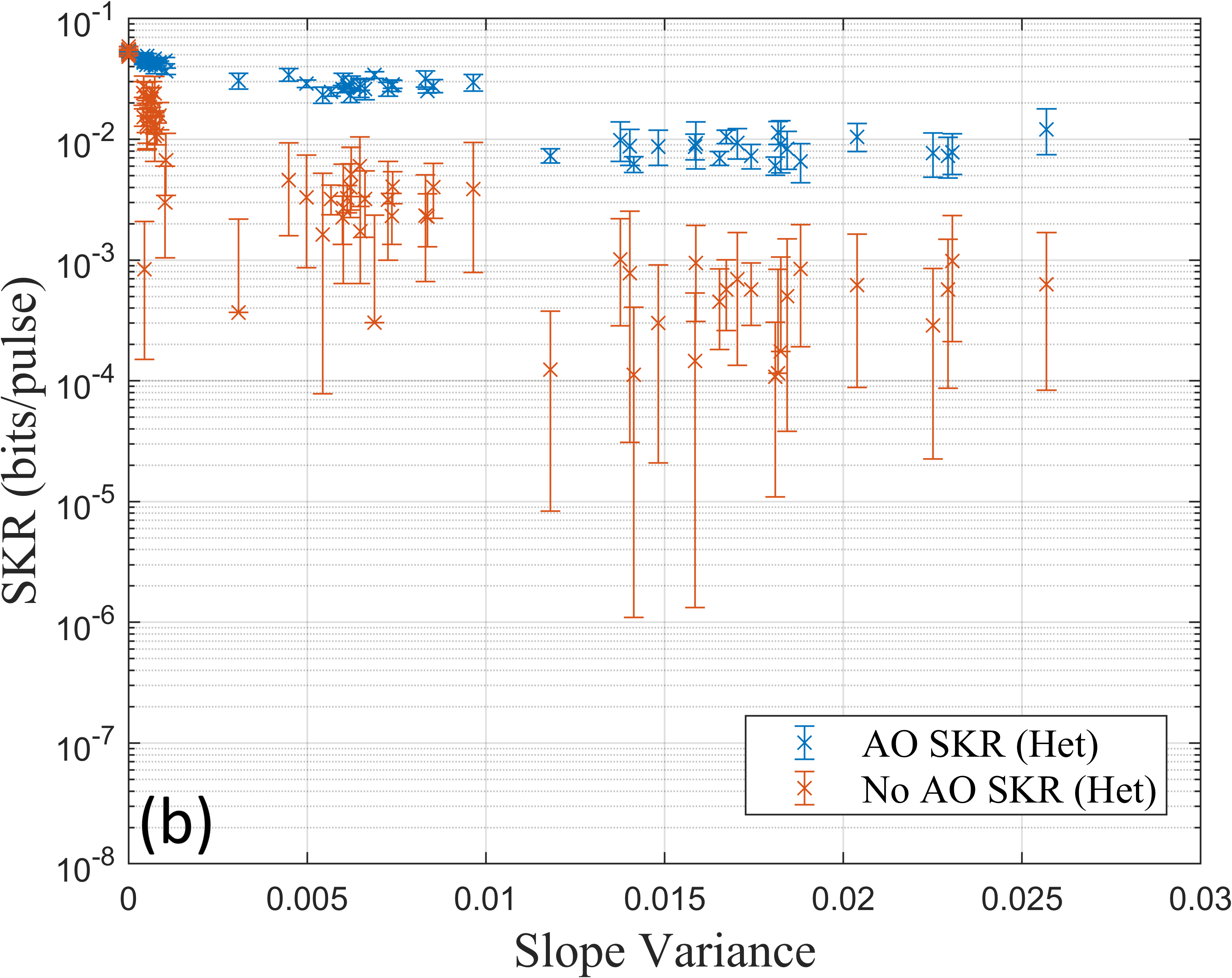}
    \caption{SKRs for Gaussian modulated CVQKD in the 60~m channel both with and without AO using  (a) homodyne and (b) heterodyne detection based on the measured visibilities and slope variances. Negative SKR values due to a small visibility have been omitted.}
    \label{fig:SKRs60}
\end{figure}

The corresponding SKRs for homodyne and heterodyne detection with and without the AO system are shown in Figure \ref{fig:SKRs60}. The calculated SKRs for heterodyne detection were included to study the effects of extending the system, which currently uses homodyne detection, to use heterodyne detection. The transmittance, $T$, before any turbulence from the heat gun was introduced was measured to be $T=0.4433$. The detector efficiency \cite{HQEPD} was $\eta_{\mathrm{det}} = 0.99$. The electronic noise was measured to be $v_{\mathrm{el}} = 0.027$ SNU using Equation \ref{eqn:electronicNoise}. The SKR was calculated using Equation \ref{eqn:SKR_asy} assuming a reconciliation efficiency of $\beta = 0.9$, which is a conservative value that produces positive SKRs \cite{bai2017high}. The modulation variance, $V_A$, was optimised for both homodyne and heterodyne detection ($V_{A, \mathrm{hom}} = 0.3$ SNU, $V_{A, \mathrm{het}} = 2$ SNU), and an excess noise value of $\xi = 0.001$ SNU was used for comparison between homodyne and heterodyne detection.

In Figure \ref{fig:SKRs60}, it can be seen that the AO system is capable of increasing the SKR for both homodyne and heterodyne detection, where the SKRs with the AO system are larger than without it. The SKR decreases when the slope variance increases. This is from the turbulence injected by the heat gun which decreases the visibility, and therefore the transmittance. The difference in the SKRs with and without the AO is greater in the heterodyne than in the homodyne case by a factor of approximately 10. This can be attributed to heterodyne detection being able to measure both quadratures simultaneously (this comes at a cost of having an extra 3 dB loss), and has been calculated using the SKR equations in combination with the reformulation of the covariance matrix (Section \ref{sec:SKRC}). In addition, the SKR with the AO system has less spread in each of the heat gun settings (Figure \ref{fig:SKRs60}). This is due to the visibility having a smaller standard deviation in each of the heat gun settings when the AO is used (Table \ref{tab:MeanSD}). It is important that the visibility has a small uncertainty to prevent larger uncertainties in the SKR due to the non-linear analytic relationship between the visibility and the SKR (Section \ref{sec:SKRC}). Note that negative SKR values from small values in the visibility have been omitted in Figure \ref{fig:SKRs60}. The AO system not only increases the SKR, but can also allow a more precise estimate of it due to a smaller error in the measured visibility; a significant improvement for CVQKD in free-space channels where the transmittance fluctuates due to turbulence.

\subsection{The 30 m Free-Space Channel}
\begin{figure}[htp!]
    \centering
    \includegraphics[width=0.7\textwidth]{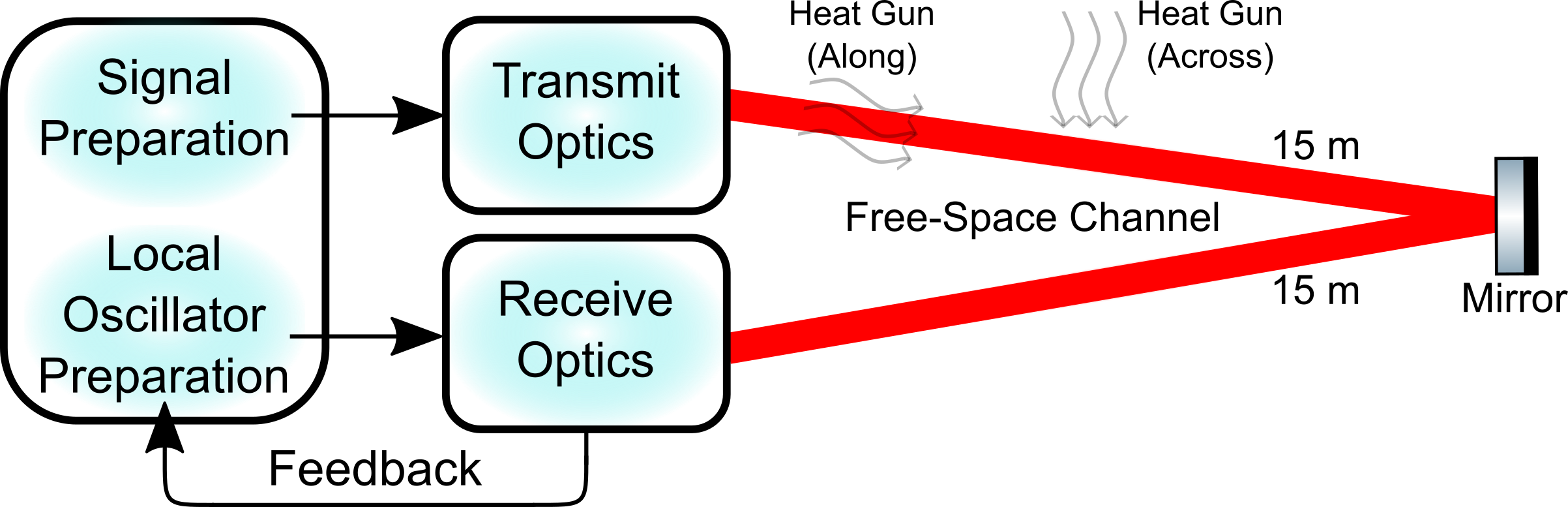}
    \caption[The 30 m free-space channel]{The 30 m free-space channel, where the laser is reflected from the Transmit Optics to the Receive Optics. Two orientations of the heat gun relative to the channel were used: across and long.}
    \label{fig:30mSetup}
\end{figure}

The 30~m channel was arranged as shown in Figure \ref{fig:30mSetup}. A mirror was used to reflect the quantum signal from the transmitter to the receiver which were placed together on an optical bench.

\begin{figure}[htp!]
    \centering
    \includegraphics[width=0.45\textwidth]{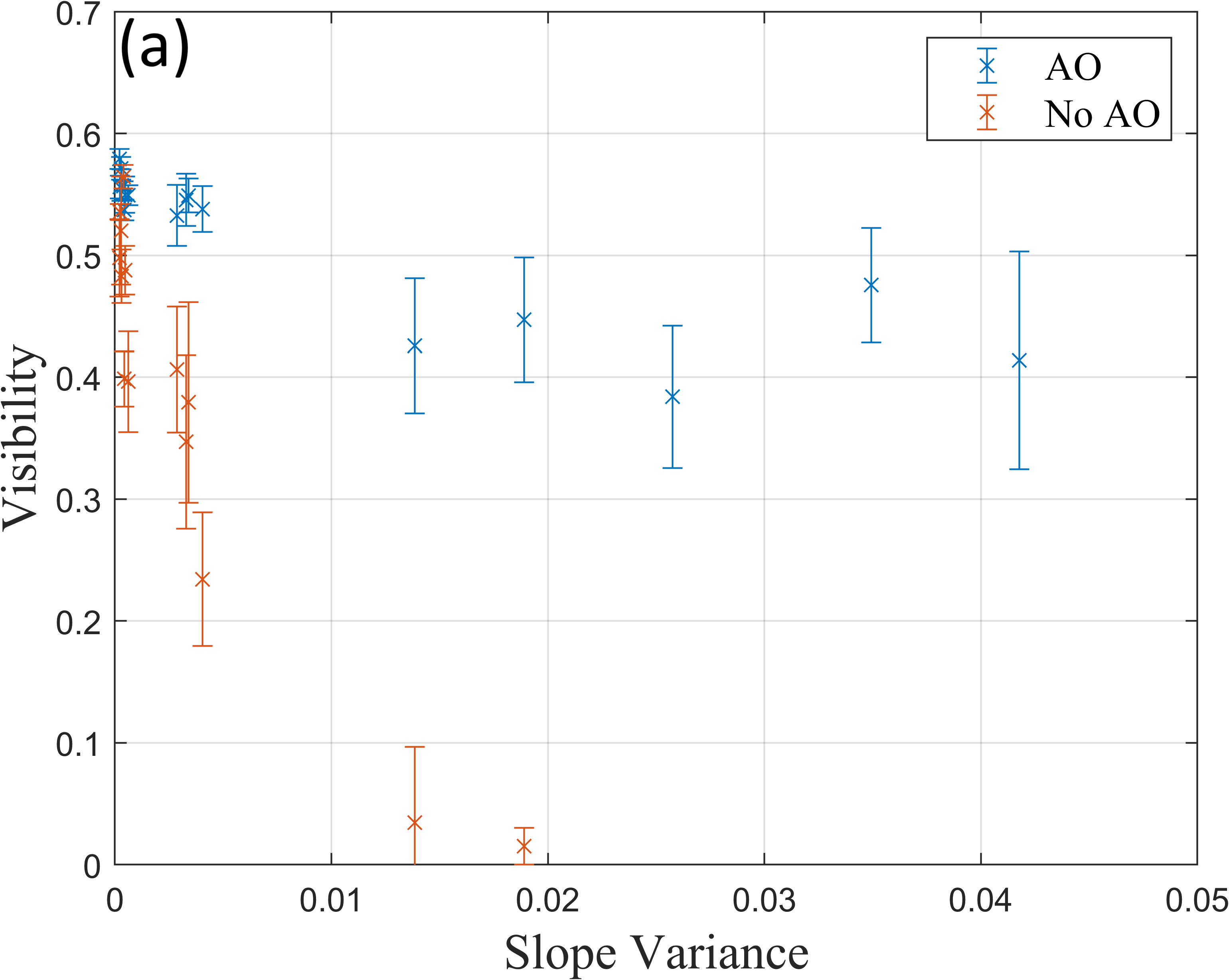}
    \includegraphics[width=0.45\textwidth]{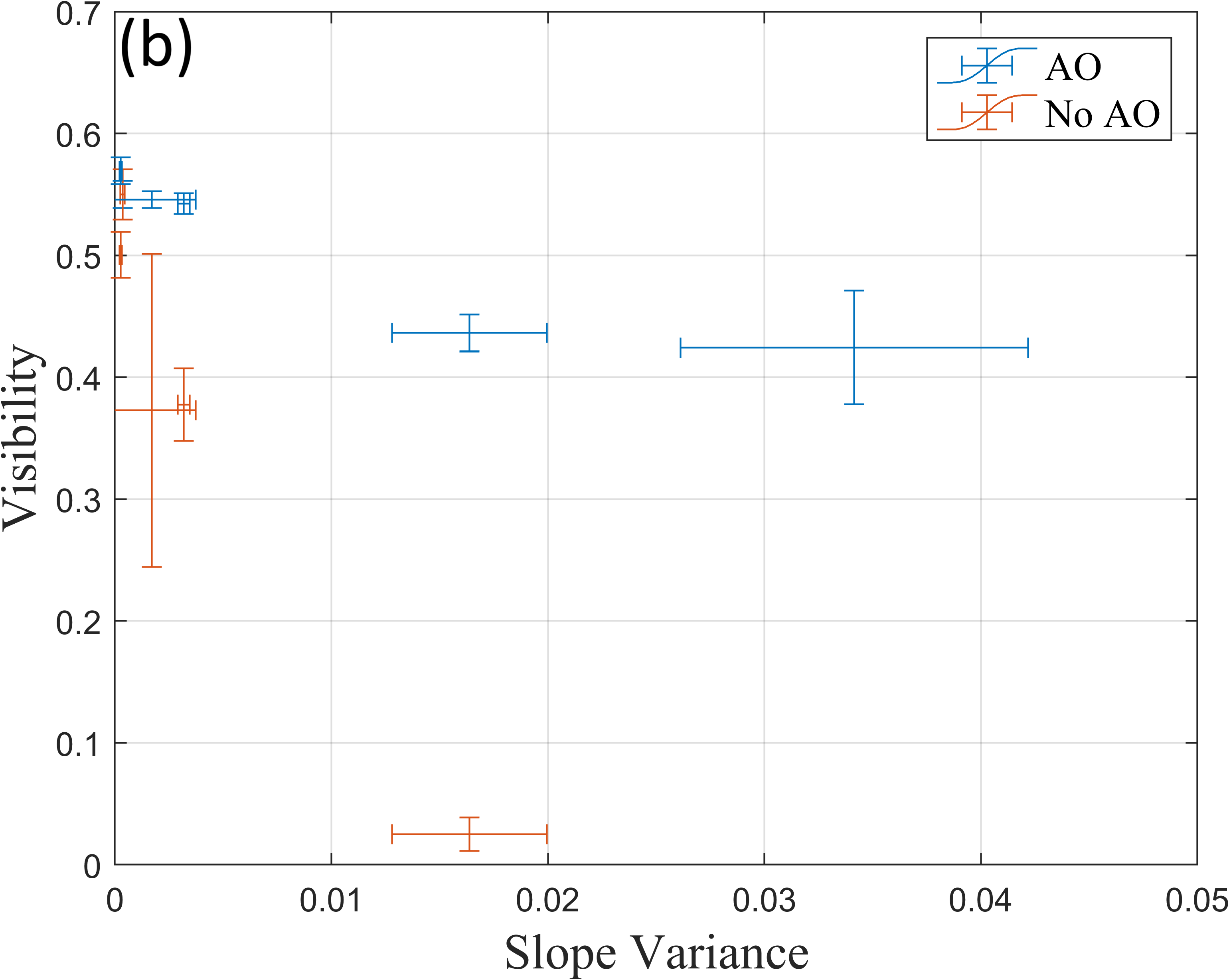}
    \caption{The (a) individual visibility measurements and (b) the mean visibility for each heat gun power setting, as a function of the slope variance both with and without the AO system in the 30 m free-space channel.}
    \label{fig:Visibility30m}
\end{figure}

The effects of turbulence in the 30 m free-space channel are displayed in Figure \ref{fig:Visibility30m}. The system had a visibility of $\sqrt{\eta_{\mathrm{vis}}} = 0.55$ when there is no turbulence from the heat gun. It was found that for the heat gun settings used in the 60~cm channel by having the hot air from the heat gun flow across the channel (in the 30~m channel, the hot air flows across both 15~m arms), smaller slope variances were measured. This resulted in the majority of the visibilities measured to be below a slope variance of $\approx$ 0.005 as the turbulence introduced was present in a small portion of the overall 30~m channel (the heat gun had a nozzle diameter of approximately 40~mm). It was decided to have the hot air flow along one of the 15~m arms (without the heat gun obstructing the beam), which resulted in larger measurements in the slope variance. This resulted in more phase-lock breaks without the AO system. The data points on Figure \ref{fig:Visibility30m} were not included as the measurements resulted in a phase drift on the oscilloscope meaning that the phase quadrature of the coherent state was not measured.

It was found that a phase-lock could be achieved at larger slope variances (Slope Variance $>$ 0.025) without the AO system and a visibility could be measured. However, the phase-lock would only last $\approx$ 2~s, after which it would break due to higher turbulence (high heat gun power setting along the channel). In addition, it would take over 3 times as long to achieve a phase-lock. These data points and the corresponding measurements with the AO system have been labelled ``Lucky Imaging", so named because of the chance associated with taking measurements with a phase-lock (Appendix \ref{AppendixC}). In comparison, non-lucky-imaging measurements were made with an immediate phase-lock being maintained. It was found that the duration of the phase-lock decreased as the slope variance increased.

The results of the visibility measurements in the 30~m channel (Figure \ref{fig:Visibility30m}) both with and without the AO system are consistent with the results from 60~cm channel (Figure \ref{fig:Visibility60cm}). In particular, there is an immediate drop in the visibility without using the AO system and that the use of the AO system largely retrieves the lost visibility. In addition, the standard deviation in the visibility of each measured coherent state decreases when using the AO system compared to when the AO system is not used as shown in Table \ref{tab:MeanSD30}. The variance in the visibility increases with turbulence due to higher levels of scintillation, similar to the 60~cm channel. With a high power heat gun setting across the free-space channel, the measured standard deviation in the visibility stays the same ($0.0533 \pm 0.0031$ with the AO system and $0.0385 \pm 0.0332$ without the AO system). This could be the AO system failing to make corrections compared to lower heat gun settings. A visibility measurement with a stable phase-lock could not be made beyond a slope variance of approximately 0.2 without the AO system. 

\begin{table} [htp!]
\centering
\caption{Mean Standard Deviation for each Measured Coherent State}
\scalebox{1}{
\begin{tabular}{|l|c|c|}
    \bottomrule
    Heat Gun Setting &  AO &  No AO \\ \toprule \bottomrule
    Ambient (no heat gun) & $0.0073$ & $0.0130$ \\
    \hline
    Low (across) & $0.0101$ & $0.0327$ \\
    \hline
    Medium (across) & $0.0201$ & $0.0684$\\
    \hline
    High (across) & $0.0533$ & $0.0385$\\
    \hline \hline
    Low (along) & $0.0139$ & $0.0388$\\
    \hline
    Medium (along) & $0.0649$ & N/A\\
    \hline
    \toprule
\end{tabular}}
\label{tab:MeanSD30}
\end{table}

The mean visibilities with and without the AO system at each heat gun setting, along with the retrieved visibility (difference) are shown in Table \ref{tab:MeanVisibilities30} and plotted in Figure \ref{fig:Visibility30m}b. The results show that with the hot air flow across the channel, the highest retrieved visibility occurred with the high heat gun setting. When the hot air flow was along the channel, the highest retrieved visibility occurred with the medium heat gun setting when a phase-lock could not be maintained without the AO system. This shows that the AO system can both increase the visibility and correct the distorted beam to achieve and maintain a phase-lock when a phase-lock cannot be achieved without it.

\begin{table} [htp!]
\centering
\caption{Mean Visibility}
\scalebox{1}{
\begin{tabular}{|l|c|c|c|}
    \bottomrule
    Heat Gun Setting &  AO &  No AO & Difference \\ \toprule \bottomrule
    Ambient (no heat gun) & $0.550$ & $0.550$ & $0$ \\
    \hline
    Low (across) & $0.5694$ & $0.5004$ & $0.0690$\\
    \hline
    Medium (across) & $0.5424$  &  $0.3775$ & $0.1649$ \\
    \hline
    High (across)& $0.4364$  & $0.0249$ & $0.4115$\\
    \hline \hline
    Low (along) & $0.5457$  & $0.3728$ & $0.1729$ \\
    \hline
    Medium (along) & $0.4244$ & N/A & $0.4244$ \\
    \hline
    \toprule
\end{tabular}}
\label{tab:MeanVisibilities30}
\end{table}

\begin{figure}[htp!]
    \centering
    \includegraphics[width=0.45\textwidth]{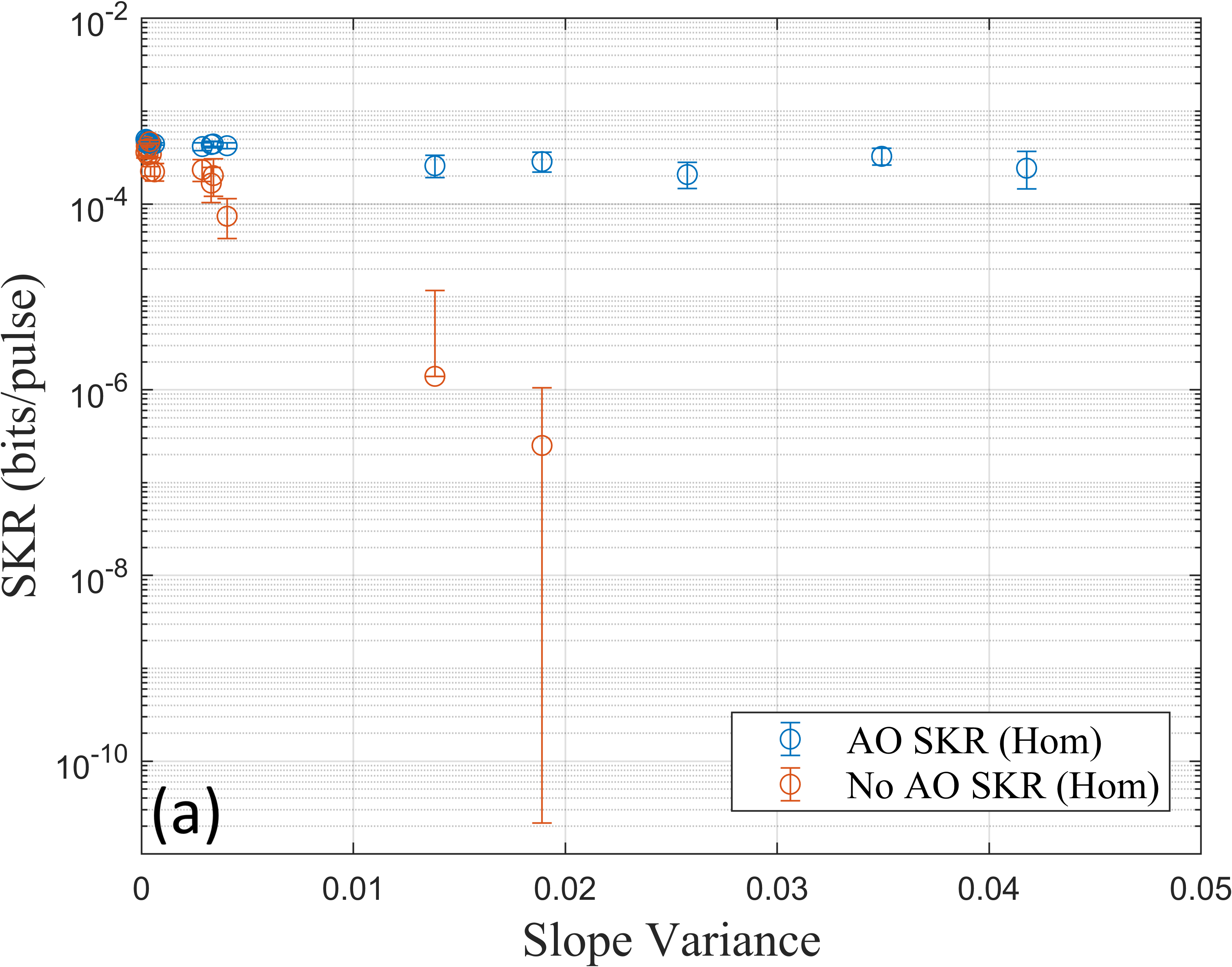}
    \includegraphics[width=0.45\textwidth]{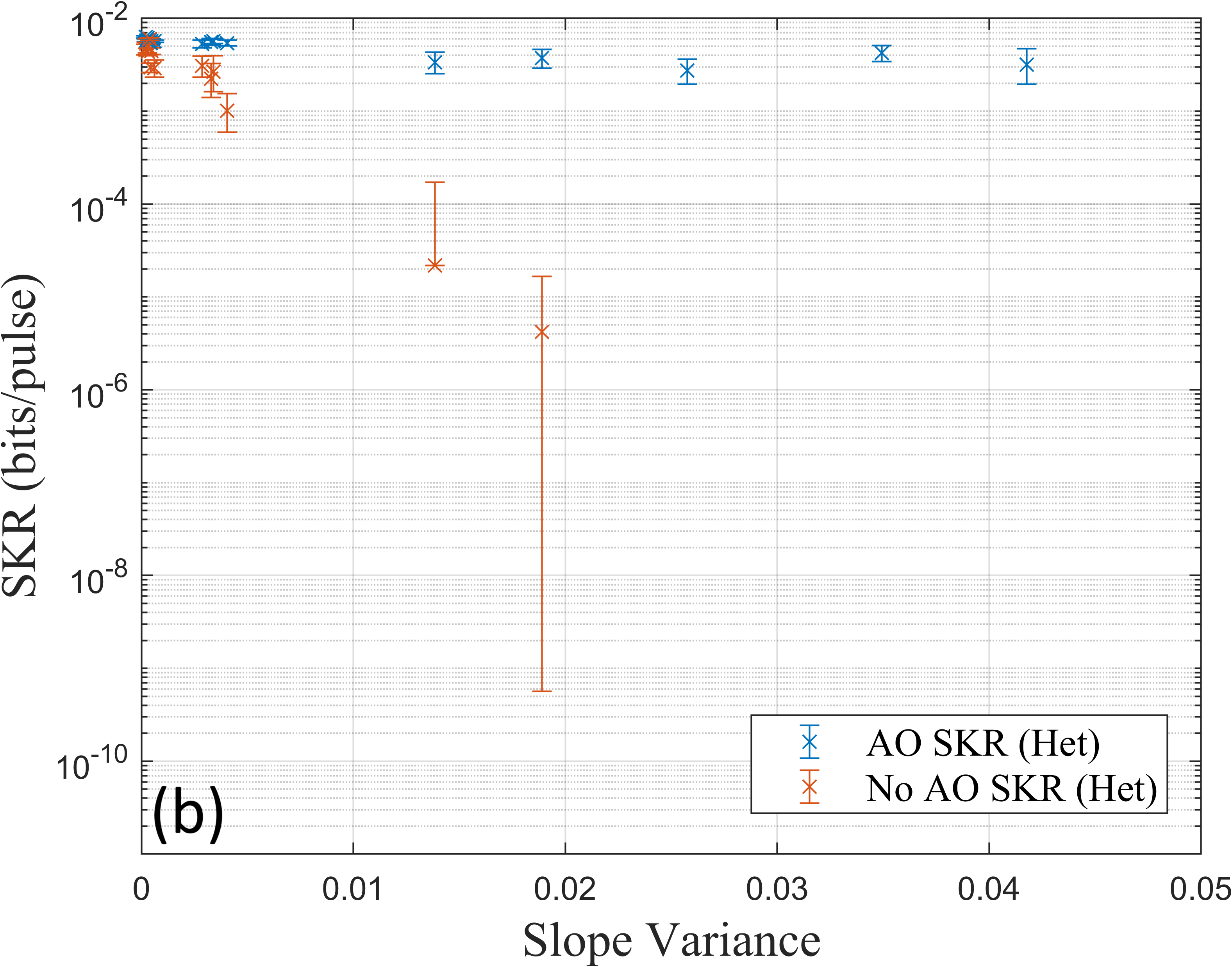}
    \caption{Calculated SKRs for Gaussian modulated CVQKD in the 30~m channel with and without AO using  (a) homodyne and (b) heterodyne detection.}
    \label{fig:SKRs30}
\end{figure}

The corresponding SKRs (Equation \ref{eqn:SKR_asy}) for homodyne and heterodyne detection, both with and without the AO system, are shown in Figure \ref{fig:SKRs30}. The same parameters were used as for the 60~cm channel. However, the transmittance without turbulence from the heat gun was measured to be $T=0.0644$, resulting in smaller overall SKRs. The results show that the AO system in the 30~m channel is also capable of increasing the SKR and decreasing its spread for both homodyne and heterodyne detection, where the SKRs with heterodyne detection are greater by a factor of approximately 10. 

The visibility variation, represented as standard deviation error bars, of some coherent states without the AO system in the 60~cm and 30~m channels had values below zero as shown in Figure \ref{fig:Visibility60cm} and Figure \ref{fig:Visibility30m}. This is not physically possible. Consequently, the calculated corresponding SKRs for the 60~cm (Figure \ref{fig:SKRs60}) and the 30~m channel (Figure \ref{fig:SKRs30}) have been omitted.

\subsection{Noise}
\begin{figure}[htp!]
    \centering
    \includegraphics[width=\textwidth]{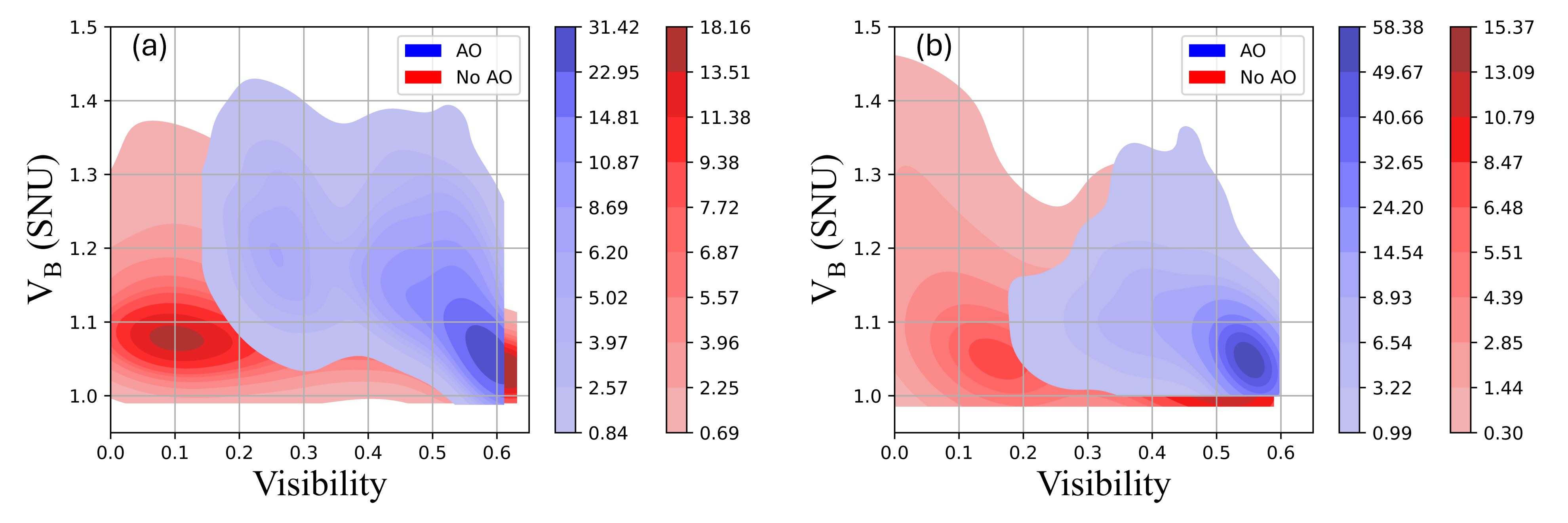}
    \caption{Density distribution for coherent state variance and visibility with and without the AO system in the (a) 60 cm and (b) 30 m free-space channel.}
    \label{fig:VarVisKDE60}
\end{figure}

The density distribution of the measured coherent state variance is plotted with and without the AO system for the 60 cm and 30 m channel as shown in Figure \ref{fig:VarVisKDE60}. It can be seen that the AO distribution is slightly above the No AO distribution by up to 0.1 SNU and 0.05 SNU in the 60 cm and 30 m channels, respectively. This indicates that the AO injects noise when used. This occurs for medium and high heat gun power settings and may come from the more vigorous corrections (more movement in the actuators in the deformable mirror for corrections) made by the AO system.

It was observed that in the 30 m channel ($T = 0.0644$) the light on the WFS was less intense than in the 60~cm channel ($T = 0.4433$). This could lead to less responsive closed-loop feedback due to a lower amplitude input signal, more relaxed corrections (less movement in the actuators in the deformable mirror for corrections), and therefore less injected noise. This is shown in the lower levels of $V_B$ with the AO system in the 30 m channel (Figure \ref{fig:VarVisKDE60}b) compared to the 60 cm channel (Figure \ref{fig:VarVisKDE60}a) when the visibility is above approximately 0.15. The large values of $V_B$ (up to 1.45 SNU) around a visibility of zero (Figure \ref{fig:VarVisKDE60}b) could be attributed to the measured coherent states with a large variance from instantaneous phase-lock breaks due to small visibility.

Some coherent states have a measured variance, $V_B$, below 1 SNU (Figure \ref{fig:VarVisKDE60}). This could be the consequence of unintended squashing where the variance falls below the standard quantum limit in one quadrature \cite{Wiseman1999Squashed}; in this case, the phase quadrature. This comes from the use of a feedback loop to adjust the phase of the LO  for phase-locking \cite{Wiseman1999Squashed}.

The SKRs were not calculated based on the excess noise from Bob's measured variance. This is because the excess noise, calculated using Equation \ref{eqn:excessnoise} for the majority of coherent states in Figure \ref{fig:VarVisKDE60}a (60~cm channel) and Figure \ref{fig:VarVisKDE60}b (30~m channel) were above the maximum tolerable excess noise to produce positive SKRs based on the transmittance without turbulence from the heat gun \cite{Denys2021Explicit}. 

\begin{figure}[htp!]
    \centering
    \includegraphics[width=0.55\textwidth]{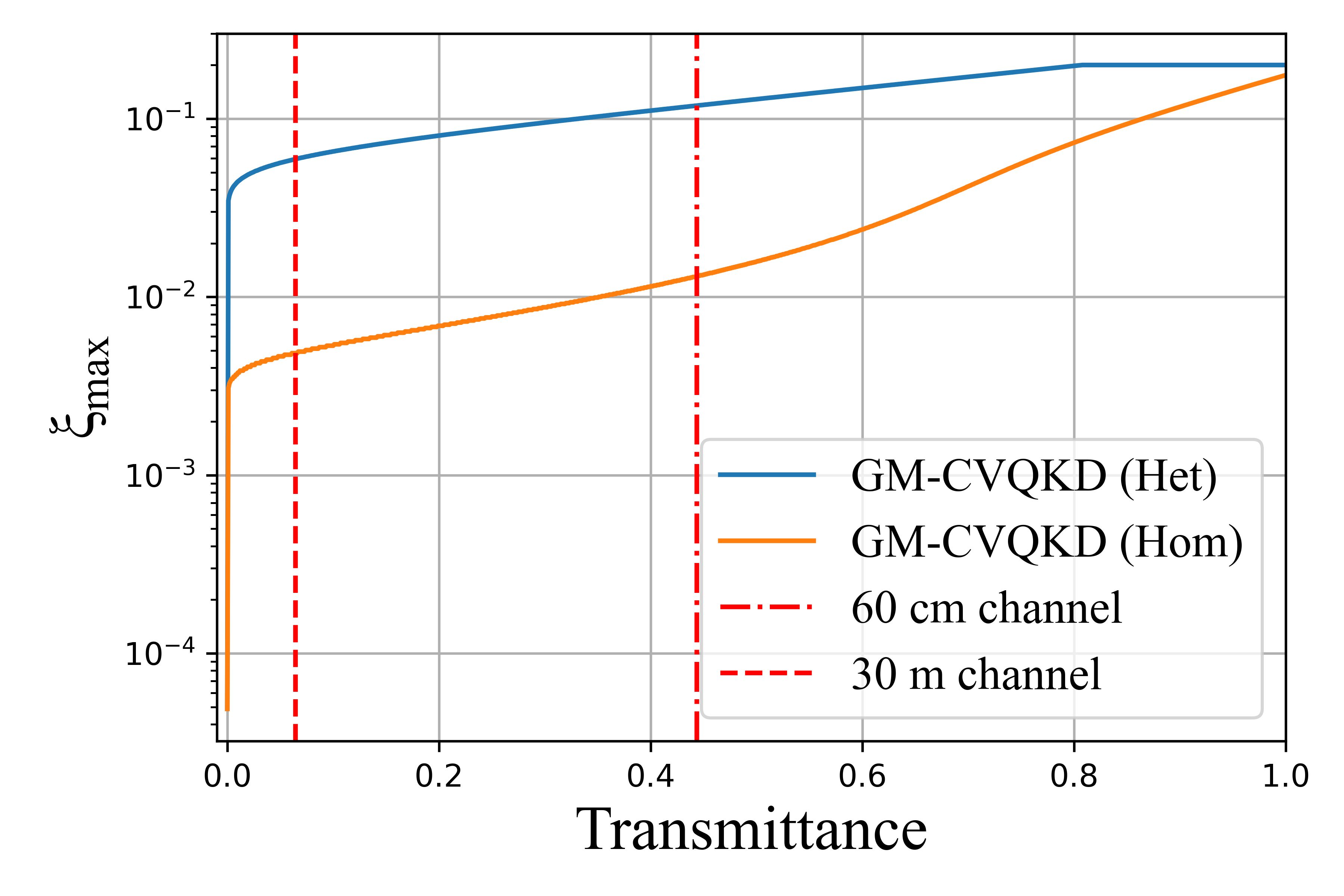}
    \caption{Maximum excess noise allowed for GM-CVQKD using heterodyne and homodyne detection. $\beta = 0.9$. The vertical lines show the transmittance of the 60~cm and 30~m channels without turbulence from the heat gun. The graphs were produced using the process in Ref. \citenum{Denys2021Explicit}.}
    \label{fig:maximalexcessnoise}
\end{figure}

Figure \ref{fig:maximalexcessnoise} shows that the maximum excess noise allowed for the 60~cm and 30~m channel are approximately 0.005 SNU and 0.011 SNU, respectively, using homodyne detection, assuming $\beta = 0.9$ and an optimised $V_A$. For heterodyne detection, the maximum excess noise allowed for the 60~cm and 30~m channel are approximately 0.006 SNU and 0.11 SNU respectively. Hence, the chosen excess noise value of $\xi = 0.001$ SNU for the calculation of the SKR based on the visibility (Figure \ref{fig:SKRs60} and Figure \ref{fig:SKRs30}) which is below the maximum tolerable excess noise. Given that the transmittance will only decrease when turbulence increases, the maximum tolerable excess noise will further decrease.

\section{Discussion}

The results show that the use of the AO system yields larger visibilities, translating to higher and more positive SKRs. In addition, the variance in measured visibilities decreases, leading to more precise estimates of the SKR. However, this comes at the price of more injected noise from the AO system. The additional excess noise introduced when using the AO system can be attributed to the closed-loop bandwidth at 135 Hz of the system which is below the operating frequency of 2~kHz \cite{bennet2024NGTF}. It is crucial that this source of noise be mitigated as increases in excess noise decrease the resulting SKR \cite{Laudenbach2018Continuous,Denys2021Explicit}. A trusted noise framework was used for the SKR analysis, where the electronic noise at Bob was trusted and consequently subtracted from the measured variance at Bob. In addition to this, the excess noise from different parts of the CVQKD system can be characterised, trusted, and subtracted from the measured variance at Bob. These include, but are not limited to, excess noise during baseline measurements, injected excess noise from the AO system, excess noise from Alice during quantum signal preparation. By identifying and characterising sources of excess noise, noise in the free-space channel due to turbulence could be isolated and more accurately characterised.

The measured slope variance in both channels for coherent state measurements were restricted to regimes where phase-locking between the quantum signal and LO was maintained. For a more robust system, phase-locking must be maintained at higher slope variances ($> 0.02-0.03$). It was found that having a small quantum signal to LO power ratio, decreases the excess noise but at the cost of a fragile phase-lock. Conversely, a large power ratio increases the excess noise but maintains the phase-lock much better. A quantum signal to LO power ratio of approximately 1:1000 was used for both the 60 cm and 30 m free-space channel when there was no turbulence from the heat gun. However, it should be noted that this power ratio will vary when turbulence is introduced, as the quantum signal power will undergo different levels of attenuation. An optimum power ratio can therefore be determined. Given that turbulence decreases the visibility and therefore the overall power of the quantum signal (turbulence causes beam wandering and scintillation, displacing the beam from the photodiodes), the phase-lock becomes more fragile. By using the AO system, the visibility is recovered, and therefore the overall power of the quantum signal, strengthening the phase-lock.

The experiment could have been improved by further optimising the optical setup to increase the visibility (ideally to $\sqrt{\eta_{\mathrm{vis}}} = 1$) when there is no turbulence from the heat gun. This could be achieved through better alignment and polarisation control for better mode-matching between the signal and LO. This would lead to a better phase-lock allowing for larger turbulence strengths to be examined with and without the AO system.

The measurements of visibility based on lucky imaging without the AO system are transient. They would translate to an instantaneous positive SKR. A common practice in estimating the SKR in free-space channels where the transmittance fluctuates is to use the mean transmittance \cite{Qu2017High,Qu2018High,usenko2018stabilization}. This would mitigate the influence of the resulting instantaneous positive SKRs on the overall estimated SKR caused by favourable fluctuations in the transmittance. By using the AO system, the transmittance is increased and its fluctuations decreased. This leads to increased SKRs with more precise estimates.

The peak performance of the AO system (largest interferometric visibility recovered) does not occur at the same slope variance for the two channels. This is attributed to the different phase and amplitude aberrations introduced by the heat gun in the two channels. Given the different wavefront aberrations in the channels, the measured slope variance would be different for the same power setting on the heat gun. This was observed when the same heat gun power setting was applied across the channels, which yielded higher slope variances in the 60~cm channel than in the 30~m channel. It was also observed in the 30~m channel where having the turbulence from the heat gun along the channel yielded higher slope variances than having it flow across the channel. In future, a more standard measurement of turbulence, such as the scintillation index, would yield to more accurate characterisations of turbulence. 

The experiment can be further extended into a full-stack CVQKD system where a secret key is encoded into the coherent states and decoded once measured. Upon the development of a full-stack CVQKD system, the improvements of AO on SKRs could be further examined.

\section{Conclusions}
The detrimental effects of turbulence on the measured visibility and variance of coherent states were experimentally studied in 60 cm and 30 m turbulent free-space channels through channel characterisation. The study showed that turbulence decreases the visibility. The use of an AO system mitigates some of the reduction in visibility due to turbulence through wavefront correction. In addition, the AO system decreases the variance in the measured visibilities. This translates to increased SKRs with more precise estimation; a favourable result for free-space CVQKD.

\section*{Acknowledgements}

Mikhael Sayat is a University of Auckland Doctoral Scholar. We would like to acknowledge funding and support from the Next Generation Technologies Fund from the Defence \& Science Technology Group, Australia. We would like to acknowledge support from A*STAR under Project No. C230917009, and Q.InC Strategic Research and Translational Thrust.

\section*{Author contributions statement}

M.S., M.B., M.C., and O.T. conducted experiments and analysed results; E.J. conducted experiments; O.T., F.B., P.K.L, N.R., and J.C. provided editing and feedback. All authors reviewed the manuscript. 

\section*{Additional information}

The authors have no competing interests to declare that are relevant to the contents of this article.
\newpage
\appendix
\section{SKR Calculation}
\label{AppendixA}
The mutual information is calculated as

\begin{equation}
    I_{AB, \mathrm{hom}} = \frac{1}{2}\mathrm{log_2}\left(1 + \frac{2\eta_{\mathrm{vis}}\eta_{\mathrm{det}}T\alpha^2} {2 + \eta_{\mathrm{vis}}\eta_{\mathrm{det}}T\xi}\right)
\end{equation}
for homodyne detection, and 
\begin{equation}
    I_{AB, \mathrm{het}} = \mathrm{log_2}\left(1 + \frac{2\eta_{\mathrm{vis}}\eta_{\mathrm{det}}T\alpha^2} {2 + \eta_{\mathrm{vis}}\eta_{\mathrm{det}}T\xi}\right)
\end{equation}
for heterodyne detection.

The Holevo bound is calculated as

\begin{equation}
\label{Holevo Bound 1}
\begin{aligned}
    S_{BE} = G\left(\frac{\lambda_1 - 1}{2}\right) + G\left(\frac{\lambda_2 - 1}{2}\right) - G\left(\frac{\lambda_3 - 1}{2}\right),
\end{aligned}
\end{equation}
where $\lambda_1$ and $\lambda_2$ are the symplectic eigenvalues of the covariance matrix in Equation \ref{CM}, and $G = (x+1)\log_2(x+1) - (x)\log_2(x)$. $\lambda_3$ is calculated as

\begin{equation}
\label{lambda3_hom}
\begin{aligned}
    \lambda_{3, \mathrm{hom}} = \sqrt{V\left(V - \frac{Z^2}{V_B}\right)},
\end{aligned}
\end{equation}

for homodyne detection, and

\begin{equation}
\label{lambda3_het}
\begin{aligned}
    \lambda_{3, \mathrm{het}} = V - \frac{Z^2}{V_B + 1},
\end{aligned}
\end{equation}

for heterodyne detection.

\section{Measurement Data}
\label{AppendixB}

\textbf{60 cm Channel}

\begin{table} [htb!]
\centering
\caption{Mean Slope Variances based on Heat Gun Setting with Standard Deviations}
\scalebox{1}{
\begin{tabular}{|l|c|}
    \bottomrule
    Heat Gun Setting &  Slope Variance \\ \toprule \bottomrule
    Ambient (no heat gun) & $6.5 \times 10^{-6} \pm 3.4 \times 10^{-6}$ \\
    \hline
    Low & $6.5 \times 10^{-4} \pm 1.9 \times 10^{-4}$ \\
    \hline
    Medium & $0.0065 \pm 0.0015$ \\
    \hline
    High & $0.018 \pm 0.0035$ \\
    \toprule
\end{tabular}}
\end{table}

\begin{table} [htb!]
\centering
\caption{Mean Standard Deviation in the Visibility of Measured Coherent States (Table \ref{tab:MeanSD}) with Standard Deviations}
\scalebox{1}{
\begin{tabular}{|l|c|c|}
    \bottomrule
    Heat Gun Setting & AO & No AO \\ \toprule \bottomrule
    Ambient (no heat gun) & $0.001 \pm 0.001$ & $0.002 \pm 0.001$ \\
    \hline
    Low & $0.014 \pm 0.007 $ & $0.049 \pm 0.013$\\
    \hline
    Medium & $0.024 \pm 0.009$ & $0.056 \pm 0.024$ \\
    \hline
    High & $0.037 \pm 0.015$ & $0.037 \pm 0.012$ \\
    \toprule
\end{tabular}}
\end{table}

\begin{table} [htb!]
\centering
\caption{Mean Visibilities (Table \ref{tab:MeanInterferometricVisibilities}) with Standard Deviations}
\scalebox{1}{
\begin{tabular}{|l|c|c|c|}
    \bottomrule
    Heat Gun Setting & AO & No AO & Difference \\ \toprule \bottomrule
    Ambient (no heat gun) & $0.60 \pm 0.0054$ & $0.60 \pm 0.012$ & $0.00 \pm 0.0174$ \\
    \hline
    Low & $0.55 \pm 0.017$ & $0.33 \pm 0.091$ & $0.22 \pm 0.108$\\
    \hline
    Medium & $0.45 \pm 0.022$ & $0.15 \pm 0.041$ & $0.30 \pm 0.063$ \\
    \hline
    High & $0.26 \pm 0.024$ & $0.060 \pm 0.021$ & $0.20 \pm 0.045$ \\
    \toprule
\end{tabular}}
\end{table}

\newpage
\textbf{30 m Channel}
\begin{table} [htp!]
\centering
\caption{Mean Slope Variances based on Heat Gun Setting in the 30~m Channel with Standard Deviations}
\scalebox{1}{
\begin{tabular}{|l|c|c|}
    \bottomrule
    Turbulence Strength &  Across &  Along\\ \toprule \bottomrule
    Ambient (no heat gun) & \multicolumn{2}{c|}{$0.00035 \pm 0.00011$}\\
    \hline
    Low & $ 0.0003 \pm 0.0001$ & $0.0017 \pm 0.0020$  \\
    \hline
    Medium & $0.0032 \pm 0.0003$ & $0.0342 \pm 0.0080$ \\
    \hline
    High & $0.0164 \pm 0.0036$  & $0.0526 \pm 0.0233$ (Lucky Imaging) \\
    \toprule
\end{tabular}}
\end{table}

\begin{table} [htp!]
\centering
\caption{Mean Standard Deviation for each Measured Coherent State (Table \ref{tab:MeanSD30}) with Standard Deviations}
\scalebox{1}{
\begin{tabular}{|l|c|c|}
    \bottomrule
    Turbulence Strength &  AO &  No AO \\ \toprule \bottomrule
    Ambient (no heat gun) & $0.0073 \pm 0.0029$ & $0.0130 \pm 0.0085$ \\
    \hline
    Low (across) & $0.0101 \pm 0.0024$ & $0.0327 \pm 0.0113$ \\
    \hline
    Medium (across) & $0.0201 \pm 0.0056$ & $0.0684 \pm 0.0155$\\
    \hline
    High (across) & $0.0533 \pm 0.0031$ & $0.0385 \pm 0.0332$\\
    \hline \hline
    Low (along) & $0.0139 \pm 0.0054$ & $0.0388 \pm 0.0176$\\
    \hline
    Medium (along) & $0.0649 \pm 0.0218$ & N/A\\
    \hline
    \toprule
\end{tabular}}
\end{table}

\begin{table} [htp!]
\centering
\caption{Mean Visibilities (Table \ref{tab:MeanVisibilities30}) with Standard Deviations}
\scalebox{1}{
\begin{tabular}{|l|c|c|c|}
    \bottomrule
    Turbulence Strength &  AO &  No AO & Difference \\ \toprule \bottomrule
    Ambient (no heat gun) & $0.5500 \pm 0.0112$ & $0.5500 \pm 0.0205$ & $0.0000 \pm 0.0317 $ \\
    \hline
    Low (across) & $0.5694 \pm 0.0109$ & $0.5004 \pm 0.0188$ & $0.0690 \pm 0.0297$\\
    \hline
    Medium (across) & $0.5424 \pm 0.0086 $  &  $0.3775 \pm 0.0297$ & $0.1649 \pm 0.0383 $ \\
    \hline
    High (across)& $0.4364 \pm 0.0151$  & $0.0249 \pm 0.0138$ & $0.4115 \pm 0.0289$\\
    \hline \hline
    Low (along) & $0.5457 \pm 0.0067$  & $0.3728 \pm 0.1284$ & $0.1729 \pm 0.1351$ \\
    \hline
    Medium (along) & $0.4244 \pm 0.0467$ & N/A & $0.4244 \pm 0.0467$ \\
    \hline
    \toprule
\end{tabular}}
\end{table}

\newpage
\section{Lucky Imaging}
\label{AppendixC}

\begin{figure}[htp!]
    \centering
    \includegraphics[width=0.6\textwidth]{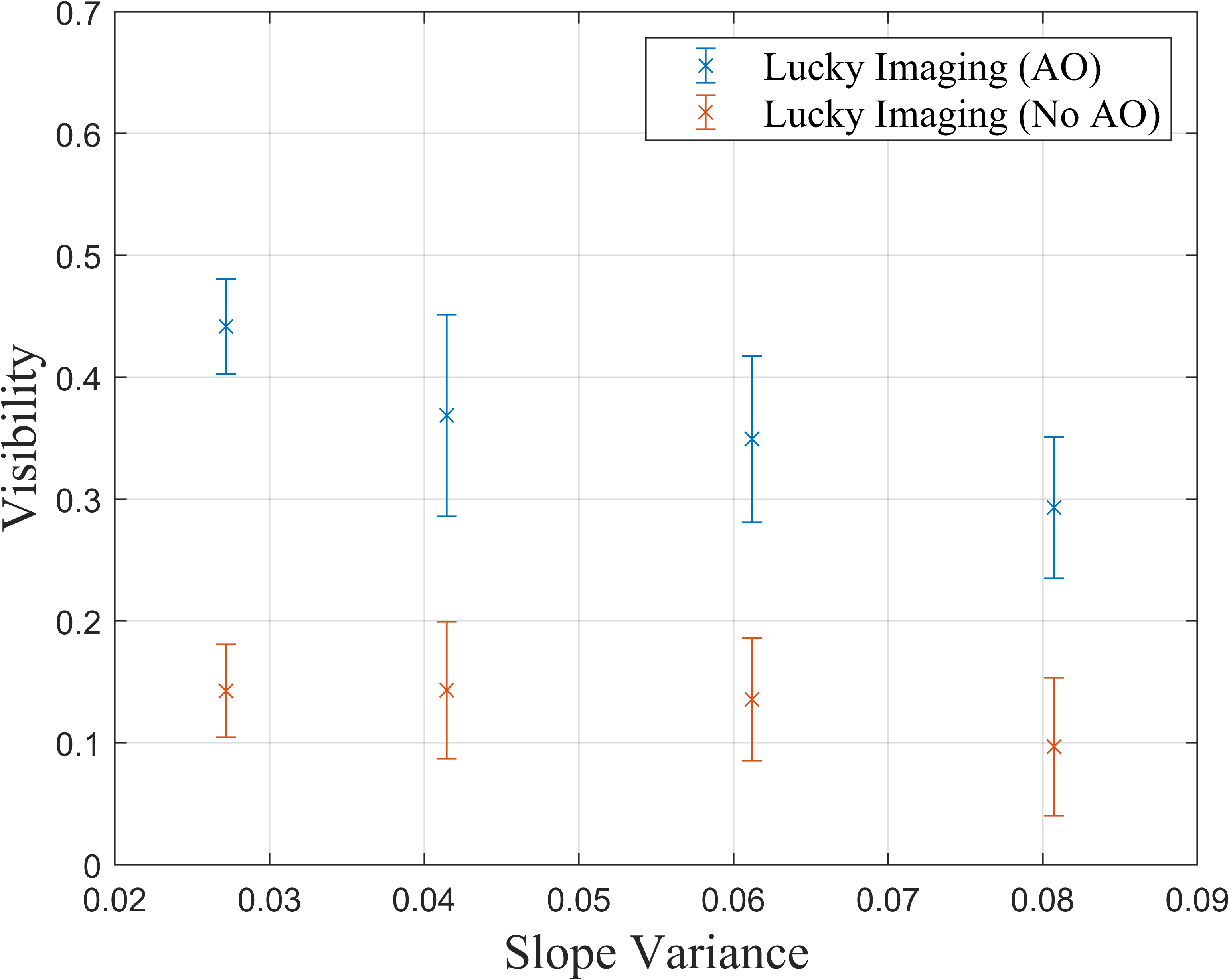}
    \caption{Visibility of `Lucky Imaging' measurements as a function of the slope variance with and without the AO system in the 30 m free-space channel.}
    \label{fig:Lucky_Imaging_VISvsSV}
\end{figure}

\newpage
\bibliography{sample}

\end{document}